\documentstyle[astrobib]{mnv2}

\input psfig
\title[Probing the dynamics of cluster-lenses]
{Probing the dynamics of cluster-lenses}

\author[Priyamvada Natarajan \& Jean-Paul Kneib]
  {Priyamvada Natarajan \& Jean-Paul Kneib\thanks{Present address:
  Observatoire Midi-Pyrenees, 14 Av. E.Belin, 31400 Toulouse, France} \\
  Institute of Astronomy, Madingley Road, Cambridge CB3 0HA}

\begin{document}
\label{firstpage}
\maketitle

\begin{abstract}
We propose a new approach to study the dynamical implications of mass
models of clusters for the velocity structure of galaxies in the
core. Strong and weak lensing data are used to
construct the total mass profile of the cluster, which is used in
conjunction with the optical galaxy data to solve in detail for the
nature of galaxy orbits and the velocity anisotropy in the central 
regions. We also examine other observationally and physically motivated 
mass models, specifically those obtained from X-ray observations and 
N-body simulations. The aim of this analysis is to understand qualitatively
the structure of the core and test some of the key
assumptions of the standard picture of cluster formation regarding
relaxation, virialization and equilibrium. This technique is applied
to the cluster Abell 2218, where we find evidence for an anisotropic
core, which we interpret to indicate the existence of a dynamically 
disturbed central region. We find that the requirement of physically 
meaningful solutions for the velocity anisotropy places stringent
bounds on the slope of cluster density profiles in the inner regions. 
\end{abstract}


\section{Introduction}

Studying the velocity structure of the cores of clusters of
galaxies promises to provide new insights into the physics of
the formation of clusters. The crucial physical consequence
of the cumulative dynamical history of a cluster is its underlying 
mass distribution. We propose a new approach to study the dynamics of
cluster galaxies using the mass profile measured from gravitational
lensing. Lensing provides the most accurate determination of the mass
profile, and is independent of assumptions as to the kinematics of the
cluster.

The dramatic arcs and multiple images produced by rich clusters
tightly constrains the mass of the cluster in the inner-most 
regions, on the scale of the Einstein radius, with typical range
$30\,<\,r_E\,<\,200\,h^{-1}_{50}$ kpc \cite{kneibsoucail95}.
Current progress in observational techniques have made 
it possible to map the cluster mass out to large radii from the weak 
shearing of faint background galaxies. While uncertainties arise due 
to the correction for the point-spread function and the 
unknown redshift-distribution of background galaxies, reliable mass 
profiles for an increasing number of clusters should be available
in the near future \cite{kaiser95b}.
Combining this with knowledge of the spatial distribution and
line-of-sight component of the velocities of cluster galaxies, and
assuming them to be good tracers of the cluster
potential well, we can solve in detail for the variation of the
velocity anisotropy parameter $\beta$ with distance from the cluster
centre. 

It is an observational challenge to perform a comprehensive redshift
survey of distant clusters, $z\,>\,0.1$ in sufficient detail to
interpret the velocity histogram and efficiently disentangle the
effects of substructure, existence of velocity anisotropy and
axisymmetric infall. However, securing 200 to 800 velocities for distant
cluster galaxies is now becoming possible with the newly developed
MOS multi-object imaging spectrographs (\citeN{yee96} and
\citeN{lefevre94}). This would enable constructing secure
line-of-sight velocity dispersion profiles for high redshift clusters.

Recent optical surveys by \citeN{colless95} and X-ray studies of
clusters by \citeN{briel95} seem to indicate that while the X-ray
isophotes of cluster cores have an overall smooth appearance most of 
them are not only dynamically young but also quite disturbed. The
nature of orbits is therefore an important indicator of the dynamical 
state of both the inner and outer regions of the cluster. Principally 
radial orbits, expected at the outskirts, are the signature of a
region dominated by infall, whereas isotropic orbits imply the
existence of a well-mixed region. 

Previous studies of the velocity dispersion profile and estimates of the 
degree of anisotropy in clusters have provided
ambiguous results, primarily due to the lack of knowledge
of the underlying mass distribution. The velocity structure of 
galaxies in clusters have been studied in detail (Coma and A2670) by 
\citeN{kent82}; \citeN{the86}; \citeN{merritt87}; \citeN{sharples88}; 
\citeN{colless95} and several other groups. 
These analyses used the observed galaxy positions
and velocities to constrain the distribution of total
mass and simultaneously find consistent and physically meaningful 
solutions for the velocity anisotropy. 

\citeN{the86} examined the uncertainties in the virial mass 
profiles derived for Coma from observational data. They showed 
that a wide range of mass models are consistent, consequently permitting a large
range of orbital structures. Mass models that were more compact 
(but had low overall masses) implied circular orbits for the galaxies 
whereas the higher mass models implied predominantly radial orbits.
\citeN{merritt87} examined the relative distribution of the dark and luminous
matter in Coma and showed that it was impossible to distinguish between
models where the galaxies trace the total mass and were on isotropic
orbits versus those in which the dark matter was very concentrated
and the galaxies were on primarily transverse orbits. Therefore, 
sufficient constraints on the total mass distribution and the velocity
anisotropy cannot be obtained simultaneously using only
the luminous tracers. In the recent survey of the Coma
cluster by \citeN{colless95}, they find that the velocity distribution
is highly non-Gaussian. The dynamics can be better interpreted in terms
of an on-going merger between two sub-clusters, thus
indicating that the system is not in virial equilibrium.

Similar studies of the cluster A2670 have also been inconclusive;
in the faint photometric and spectroscopic survey 
by \citeN{sharples88}, both extremely anisotropic models and 
nearly isotropic ones were indistinguishable in terms of goodness
of fit with respect to the available data. 

In these clusters and others that have been studied, the principal
uncertainty in the determination of the anisotropy arises from
ignorance of the distribution of the total mass. In our analysis, 
independent data from lensing that
constrains the overall mass distribution allows the elimination of
this largest source of uncertainty.

The plan of this paper is as follows: in Section 2 we review
observational probes of the internal dynamics of clusters and
construct mass models from X-ray data, lensing data, and from N-body 
simulations. The mathematical formalism that forms the
basis of our approach is presented in Sections 3 \& 4. The 
robustness of the method is demonstrated
for several fiducial forms of the total mass profile (Section 5), and 
the technique is then applied to the cluster A2218 (Section 6). 
Finally, we discuss our results and their implications
for the physical state of the core of A2218. 
Throughout this paper, we assume $H_{0} = 50\,$km s$^{-1}$Mpc$^{-1}$,
$\Omega = 1$ and $\Lambda = 0$.

\section{Observed properties of the core}

\subsection{\bf Photometric and spectroscopic studies}

With the rapid progress in photometric and spectroscopic techniques,
the precision of observationally determined parameters for clusters 
has improved significantly in the past decade. The main limitations 
are the errors incurred in establishing cluster membership
\cite{alfonso93} due to contamination from the field population for
photometrically selected samples. On the spectroscopic front,
while surveys are pushing down to fainter magnitude limits, a
commensurate gain in the number of cluster galaxies sampled 
cannot be achieved since the number of faint background field galaxies
grows more rapidly. In the following subsections we examine the
observationally determined quantities that we use in our present analysis.
\subsubsection{\bf Galaxy surface density profiles from optical data}

Galaxy surface density profiles are generally determined by fitting the
observed number density in a given cluster to physically
motivated functional forms \cite{kent82}. 
Since to define cluster membership, one usually uses the
color-magnitude relation of the E \& S0 galaxies 
(given the limited spectroscopic samples available), the computed
galaxy distribution profile preferentially probes the E \& S0 cluster 
galaxies. The other morphological types in the cluster,
could in principle have different profiles, but given the
limitations of the available data we implicitly assume that the E \&
S0's efficiently trace the overall mass distribution.

While it can be argued that there is insufficient observational
evidence for a core in the galaxy distribution in clusters (\citeN{carlberg96}
and \citeN{merritt94}), it is nevertheless instructive to examine the modified
Hubble law profile which provides a reasonable fit to sparser samples.\\
PROFILE A: \\
\begin{equation}
{\nu_{g}} (r) = {\frac {\nu_{0}}{(1 + {\frac
{r^{2}}{r_{g}^{2}}})^{1.5}}}.
\end{equation}
The core radius $r_g$ for typical clusters ranges from $150\,
h_{50}^{-1}\,$kpc to about $300h_{50}^{-1}\,$kpc. A
least squares method is then used to determine the values of
$\rho_{0}$ and core radius that simultaneously provide the best-fit to
the data. For well-sampled clusters, (\citeN{esokey95} and \citeN{cnoc96}) 
it is found that the number density of galaxies can be fit by a
generic profile of the form,\\
PROFILE B:\\
\begin{eqnarray}
{\nu_{g}} (r) = {\frac {\nu_{0}} {{({r \over s})^{\alpha}}{(1 +
{r \over s})^{2-\alpha}}}},
\end{eqnarray}
with $\alpha = 1$ and $s$ being a scale-radius ranging from
$200\,-\,400h_{50}^{-1}\,$kpc. This profile looks asymptotically like
the modified Hubble law, but has a 
central cusp. We examine both these profiles and their implications
for the core dynamics in detail in Section 5 of this paper. 

In principle, non-parametric maximum-likelihood and regression 
techniques developed by \citeN{merritt95a} and \citeN{merritt94} are more
accurate in terms of characterizing the surface number density
distribution; given the quality of data available at present for 
lensing clusters (at moderately high redshift) with a secure mass 
model, good photometry and sufficient
velocities, we restrict our current analysis to a parametric approach,
primarily on the basis of statistical adequacy given the sample sizes
that we are dealing with and the convenience of working with analytic
forms. Besides, the non-parametric methods necessarily involve
smoothing the data in order to yield confidence limits, which introduces
a bias that increases with the degree of smoothing employed and is
hence undesirable for sparse samples. For a more complete data-set 
however, a non-parametric likelihood method would be more appropriate.
 
\subsubsection{\bf Line-of-sight velocity dispersion profiles}

The only measured component of the velocity dispersion of the galaxies 
is the projection along the line-of-sight. The interpretation of this
measurement might be 
severely affected by substructure in the cluster and the presence of 
interacting sub-groups. For any
distant cluster $(\,z>\,0.1)$, the line-of-sight velocity dispersion is not
determined accurately enough at present to construct a secure radial profile.
Even for nearby clusters, the rich region close to the centre 
and the outer regions, are not sampled adequately to 
infer conclusively the asymptotic behavior. For the well-studied
cluster 
Abell 2218, the error bars in the measurements (see
Fig. 6) are too large to perform any sensible fit to the available
data and extract a radial profile.    
\subsubsection{\bf X-ray observations}

X-ray observations by Einstein, ROSAT and ASCA map the thermal bremsstrahlung emission
from the hot intracluster gas at temperatures around $10^{7}\,-10^{8}K$; 
the typical observed X-ray luminosities
range from $10^{42}\,-\,10^{45}\,$ergs s$^{-1}$ (in the 0.5 -- 4.5 keV
band). The measured gas mass within the central few Mpc is of the
order of $10^{14}\,M_\odot$, and the inferred cooling times for the gas
in some clusters is of the
order of $10^{9}$ years (\citeNP{fabian82}). The surface brightness 
profiles of clusters are sharply peaked at the centre. From X-ray
observations, with a few exceptions, 
most cluster cores appear smooth and uniform with very regular
isophotes. Accurate measurements
of the temperature profile in the cores are only now becoming 
feasible (for $z\,<\,0.1$ clusters) with the ASCA satellite. 
\subsubsection{\bf Mass profiles from X-ray data}

Standard deprojection analysis of the X-ray surface brightness profile
gives the density profile of the gas, and the cluster potential
within which the gas is confined \cite{sarazin88}. The main limiting 
assumptions are spherical symmetry, hydrostatic equilibrium, the inability to take
sub-structure into account and assumptions regarding the unknown
radial temperature profile of the gas \cite{nulsen95}. The latter assumption is
particularly important when modelling the inner parts of cluster
cores with either strong cooling flows, where one expects to have a
multiphase ICM \cite{allen95}, or in clusters which have undergone
recent mergers, resulting in a complex temperature structure. 
Preliminary results of measurements of the temperature structure of
the core of A2256 \cite{briel95} seem to indicate that this cluster,
which was believed to be one of the smoothest and most uniform from
its X-ray image shows strong evidence for temperature gradients in 
the core (\citeNP{mush96}).

\subsection{\bf N-body simulations}

N-body simulations provide the crucial link in understanding how the observed
structure in clusters arises in the context of their evolution from the initial
perturbations in the gravitational instability picture. 

High-resolution simulations that incorporate gas dynamics and
some of the important gas physics like shocks and radiative cooling
are being used to study the formation, dynamics and evolution of galaxy
clusters in the scenario where structure is built up hierarchically in
a universe dominated by cold dark
matter (\citeNP{evrard90};\citeNP{cen93}; \citeNP{navarro94a}; \citeNP{frenk95}). 
The evolution of both the dark matter
and the baryonic component can be tracked to within the resolution
limit.  While the core of an individual cluster cannot be resolved in
enough detail to understand relaxation processes, an ensemble of clusters can
be studied for their `average' properties \cite{navarro94b}.

The density profiles of clusters formed in these simulations are
sensitive to the underlying cosmological model, the initial conditions,
the accuracy of modelling gas dynamic processes and prescriptions for
galaxy and star formation. \citeN{efdw85} found that in 
simulations with only dark matter
particles, the slope of the density profile on cluster scales 
steepens with increasing $n$, where $n$ is the spectral index of the
scale-free, initial perturbations in an Einstein-de Sitter cosmology. 
It has been suggested by \citeN{crone94} and 
\citeN{navarro94b} that ensemble cluster properties, like abundance, 
clustering and density profiles might be
a useful discriminant of cosmological parameters.  However, properties of
currently simulated clusters are not consistent in detail with their observed
properties, primarily a reflection of the lack of understanding of the
physics of galaxy formation and the role of non-gravitational
processes coupled with the lack of knowledge of $\Omega$. Qualitatively though, the morphology of simulated clusters is
quite similar to ROSAT observations of X-ray clusters, and the physical
effects of mass segregation due to dynamical friction and luminosity
segregation seem to be borne out in the simulations - evidence to
support our naive theoretical picture of the formation of clusters.

In a recent paper, \citeN{navarro96} report the results of their
N-body + SPH simulations, wherein a `universal density profile' 
is found to be a good fit over a large range of scales for 
dark halos in standard CDM models. The halo profiles are more or less
isothermal, shallower than $r^{-2}$ near the 
central regions and steeper close to the virial radius.  
The density profile has the following form,
\begin{equation}
\rho \, (r) = {\frac {\rho_{0}}{{\frac{r}{r_{s}}} (1 + {\frac {r}{r_{s}}})^2}}
\end{equation}
where $r_{s}$ is a scale radius.  
The corresponding mass profile is given by:
\begin{equation}
M(r)\,=\,M_{0}\,[\,\,\ln(1\,+\,{r\over r_{s}})\,+\,{1\over(1\,+\,{r\over r_{s}})}\,].
\end{equation}
We examine this mass model and its consequences for the resulting
dynamics of cluster cores in section 5.

\subsection{LENSING BY CLUSTERS}

Clusters of galaxies have the optimum cross-section for lensing the
nearly isotropically distributed high redshift faint galaxy population.
Lensing by an extended mass concentration can be understood in terms of
a mathematical mapping (e.g. \citeNP{blandford87a} and \citeNP{fort94}) from
the source plane onto the image plane with the properties that
it conserves surface brightness and is achromatic. The deflections
produced are non-linear with impact parameter and therefore produce both
amplification and distortion of the background sources. There are two
important effects: the isotropic magnification and the
non-isotropic distortion. The isotropic magnification is caused by mass
interior to the beam and is pronounced in the region of the image
plane where the local surface mass
density $\Sigma$ is of the order of the critical surface mass density,
$\Sigma_{\rm crit}$, which occurs in the dense cores of rich clusters
producing multiple images and arcs. ($\Sigma_{\rm crit}$ depends on
the angular distance to the source and lens and hence on cosmological 
parameters). The anisotropic distortion of images is caused by the gradient of the
two-dimensional potential and characterizes the `weak lensing' regime, the signal being
arclets (single weakly sheared images) produced even at large distances from
the cluster centre.

A composite mass profile for a cluster can be constructed using a
variety of constraints from lensing effects over a range of scales. The
strong lensing regime constrains the total mass enclosed within the
`Einstein radius', while weak shear effects (measured statistically
from the ellipticities of the faint background galaxies) determine the slope of the
mass profile at the outskirts.
\subsubsection{\bf Constraints from Strong Lensing}

The input from observations for the mass modelling are arc positions,
the number of merging images and their parities, and the width, shape and
curvature of the arcs. They are used to determine the location of the
critical lines in the image plane which are then mapped back to the
source plane in the method developed by \shortciteN{kneib93b}. The
difference in parameters implied by each of the multiple images is then
minimized in the $\chi^{2}$ sense in the source plane.  In order to
calibrate the lens model, at least one arc redshift needs to be
measured. The usual mass profiles used in modelling the cluster mass distributions
are the cored isothermal profile (\citeNP{blandford87c}), the pseudo-isothermal
elliptical mass distribution (PIEMD) \cite{kassiola93} or a linear
combination of them. The PIEMD model has a 2-D surface mass density defined by:
\begin{equation}
\Sigma (r) = \frac {\Sigma_{0}}{\sqrt{1 + \frac {r^{2}}{r_{0}^2}}},
\end{equation}
The corresponding 3-dimensional density profile and the mass are,
\begin{equation}
\rho (r) = \frac {\rho_{o}}{{1 + \frac {r^{2}}{r_{0}^2}}},
\end{equation}
\begin{equation}
M(r) = M_{0} \, [{\frac {r}{r_{0}}} - \tan^{-1} {\frac {r}{r_{0}}}].
\end{equation}
Modelling the arcs with this profile, the normalization and the core
radius (which is a measure of the compactness of the mass distribution)
are determined. The core radius of most lensing clusters is observed to
be quite small, $30 h_{50}^{-1}\,$kpc$\,\leq\,r_{o}\, \leq 100
h_{50}^{-1}\,$ kpc.
\subsubsection{\bf Constraints from Weak Lensing}

The slope of the mass profile at large radii ($\,r\,>\,200
h_{50}^{-1}\,$kpc) is constrained by the observed weak distortion
effects. The weak shear $\gamma$ induced by the cluster on the
background images can be written for the circularly symmetric case
as:
\begin{equation}
\gamma\,\propto\,\left<D(z_s)\right>\,(\bar\Sigma (<r)\,-\,\Sigma(r)),
\end{equation}  
where $\left<D(z_s)\right>$ is the mean of the ratio of the angular distances
$D_{\rm {lens-source}}/D_{\rm {observer-source}}$ and $\bar\Sigma$ the mean surface density within
radius $r$.
The Kaiser-Squires technique \cite{kaiser93} defines a
mapping that relates the image ellipticities to
the relative mass map $\Sigma (r)$ for a cluster. 
To construct the surface mass density profile
one uses the statistic suggested by \shortciteN{fahlman94} \& \shortciteN{squires95},
\begin{equation}
\bar\Sigma(<r_1)\,-\,\bar\Sigma(r_1<r<r_2)\,=\,{2\Sigma_{crit}\over 1
- {r_1^2/r_2^2}}\,
\int^{r_2}_{r_1}\left<\epsilon_t\right>{dr \over r},
\end{equation}
where $\left<\epsilon_t\right>$ is the mean tangential component of the image 
ellipticities.
For this inversion, deep optical images
under exquisite seeing conditions of a wide field over the lensing
cluster are required.  Details of correction and compensation for the
anisotropy of the point-spread function and bad seeing conditions have
been demonstrated by \shortciteN{kaiser95a} and \citeN{bonnet95}. 
\subsubsection{\bf Constraints from the cD galaxy}

Dark matter in clusters is sharply peaked about the cluster centre
around which the lensed images are seen. In most clusters with
spectacular arcs, the centre of the brightest cluster galaxy and the
centre of the dark matter distribution as determined both from X-ray
and lensing studies seem to be coincident to within
the errors - of the order of a few arcseconds.

The central bright elliptical galaxies are often cDs with diffuse
halos extending out to beyond the Einstein radius. The orbits of the stars in
these halos trace the overall dark matter potential. The density
profiles of cD galaxies are fairly well-determined observationally
\shortcite{kneib95a} and are best fit by a difference
of 2 PIEMD models,
\begin{equation} 
\rho (r)\,=\,{\rho_{0}{r_{1}^2}{r_{cut}^2} \over ({r_{cut}^2}-{r_{1}^2})}\,[\frac{1}{r^2 + r_{1}^2} - \frac{1}{r^2 + r_{cut}^2}],
\end{equation}
where $r_{1}$ is the core radius and $r_{\rm cut}$ is the truncation
radius. The velocity dispersion profile of giant ellipticals is also
measured in a number of clusters (\citeNP{fisher95}) and is found to be,
$\sigma_{*} \sim 300 - 500 \,\rm km \, s^{-1}$.
Close to the cluster centre, the overall mass profile 
has to be consistent with the measured isotropic velocity dispersion
of stars in the cD \cite{jme95}.

\section{DYNAMICAL EQUATIONS}

We model the cluster as a collisionless system in which the individual
galaxies move under the influence of the mean gravitational field $\phi$
generated by all the constituents. The system is characterized by its
phase-space density $f(\bmath x \,, \bmath v \,,t)$, and a given configuration
of the system is specified by $f(\bmath x,\, \bmath v,\, t ) \, d^{3} \bmath
x \, d^{3} \bmath v$ -- the number of galaxies having positions in the
infinitesimal volume $d^{3} \bmath x$, with velocities in the range
$d^{3} \bmath v$.  It should be noted here that in phase-space $\bmath x$
and $\bmath v$ are independent variables and the potential is not a
function of $\bmath v$.  The density of points in phase-space
satisfies the continuity equation,
\begin{equation}
{\frac {D f}{D t}} = {\frac {\delta f}{\delta t}} + {\bmath v} \,. {\bmath \nabla} f -
{\bmath \nabla \phi} .  {\frac {\delta f}{\delta \bmath v}} = 0,
\end{equation}
which is the {\it collisionless Boltzmann equation}.
 Neglecting the explicit time derivative; taking the first
velocity moment; integrating over all possible velocities for a
spherical system, we obtain the {\it Jeans equation}:
\begin{equation}
{\frac{ d (\rho \, \sigma_{r}^{2})}{d r}} + {\frac {\rho  \, [ 2 \sigma_{r}^{2} -
(\sigma_{\theta}^{2} + \sigma_{\phi}^{2})]}{r}} = - \rho  \,
{\frac {d \phi}{d r}},
\end{equation}
where $\rho$ is the density profile and $\sigma_{i}^{2}$ are the 
components of the velocity dispersion. If, additionally, the velocities and the density are invariant
under rotations about the cluster centre we have,
\begin{equation}
\sigma_{t}^{2} \, \equiv \, \sigma_{\theta}^{2} \, = \, \sigma_{\phi}^{2}, 
\end{equation}
\begin{equation}
{\frac{ d (\rho  \, \sigma_{r}^{2})}{d r}} + {\frac{ 2 \beta \, \sigma_{r}^{2} 
\, \rho }{r}} = - \rho  \, {\frac {d \phi}{ d r}}.
\end{equation}
The velocity anisotropy parameter $\beta$ at a given point is,
\begin{equation}
\beta (r) = (1 - {\frac {\sigma_{t}^{2}}{\sigma_{r}^{2}}})
\end{equation}

\subsection{The Isotropic Jeans equation}

The isotropic Jeans equation is a special case of the more general equation
above, wherein the galaxies are on isotropic orbits. Hence $\beta = 0$, and
$\sigma_{r}^{2} \, = \sigma_{t}^{2}$,
\begin{equation}
{\frac {d \, (\rho \, \sigma^{2} \,)}{ d r}}= - {\frac {G M(r)
\rho(r)}{r^{2}}}\,=\,-{\rho\,v_c^{2}\over r},
\end{equation}
where $v_{c}(r)$ is defined to be the circular velocity. 
The solutions for the isotropic velocity dispersion are given by:
\begin{equation}
\sigma^{2} (R) \, = {G \over \rho} \int_{R}^{\infty} {M(r) \rho(r)
\, dr \over r^{2}}\,=\,{ 1 \over \rho}\int_{R}^{\infty}
\rho(r)\,v_c^{2}(r)\,{dr \over r}.
\end{equation}

The velocity dispersion of both the galaxies and the total mass can
therefore be computed given their respective density profiles and the
underlying mass distribution. We plot the solutions obtained for
a mass model of the form,
\begin{equation}
M (r) = M_{0} \, [{\frac{r}{r_{0}}} - \tan^{-1}{\frac{r}{r_{0}}}],
\end{equation}
with $r_0\,=\,50\,$kpc, $\sigma_0\,=\,1200\,$km s$^{-1}$,
and generic density profiles for the tracers (galaxies or isothermal gas) of the form,
\begin{equation}
\rho (r) = \frac {\rho_{o}}{{(1 + \frac{r^2}{r_{c}^2})}^{\alpha}}.
\end{equation}
The solutions for $\alpha =\,$1.0, 1.2 and 1.5 and for various values of
${r_{c}/r_{0}}\,=\,$1.0, 2.5 and 5.0 are plotted in
Fig. 1.  
The solutions have the following interesting properties:
\begin{enumerate}
\item the velocity dispersion falls in the centre for small core radii, 

\item the smaller the core radius of the tracer, the lower the central value of the velocity dispersion,

\item the mean value of the velocity dispersion is a weak function of $r_c$
but depends on the slope $\alpha$. 
\end{enumerate}
{\it{Therefore one can have different mean values of the velocity
dispersion for different components if they do not have the same radial profile.}}
 
In Fig. 1, we also plot the ratio of the velocity dispersions of dark matter and galaxies,
which provides a qualitative understanding of the velocity
bias  (as found in the numerical simulations by \citeNP{carlberg94}).   
The difference in the asymptotic slope of the density profiles of dark
matter and galaxies, and the ratio of the core radii is found to
determine the velocity bias. To first order, the asymptotic behavior
(regardless of core size) is:
\begin{equation}
\left({\sigma_{DM}\over\sigma_g}\right)^2\simeq
        {\alpha_{DM}\over\alpha_g}
\end{equation} 
 
\begin{figure*}
\centerline{\psfig{figure= 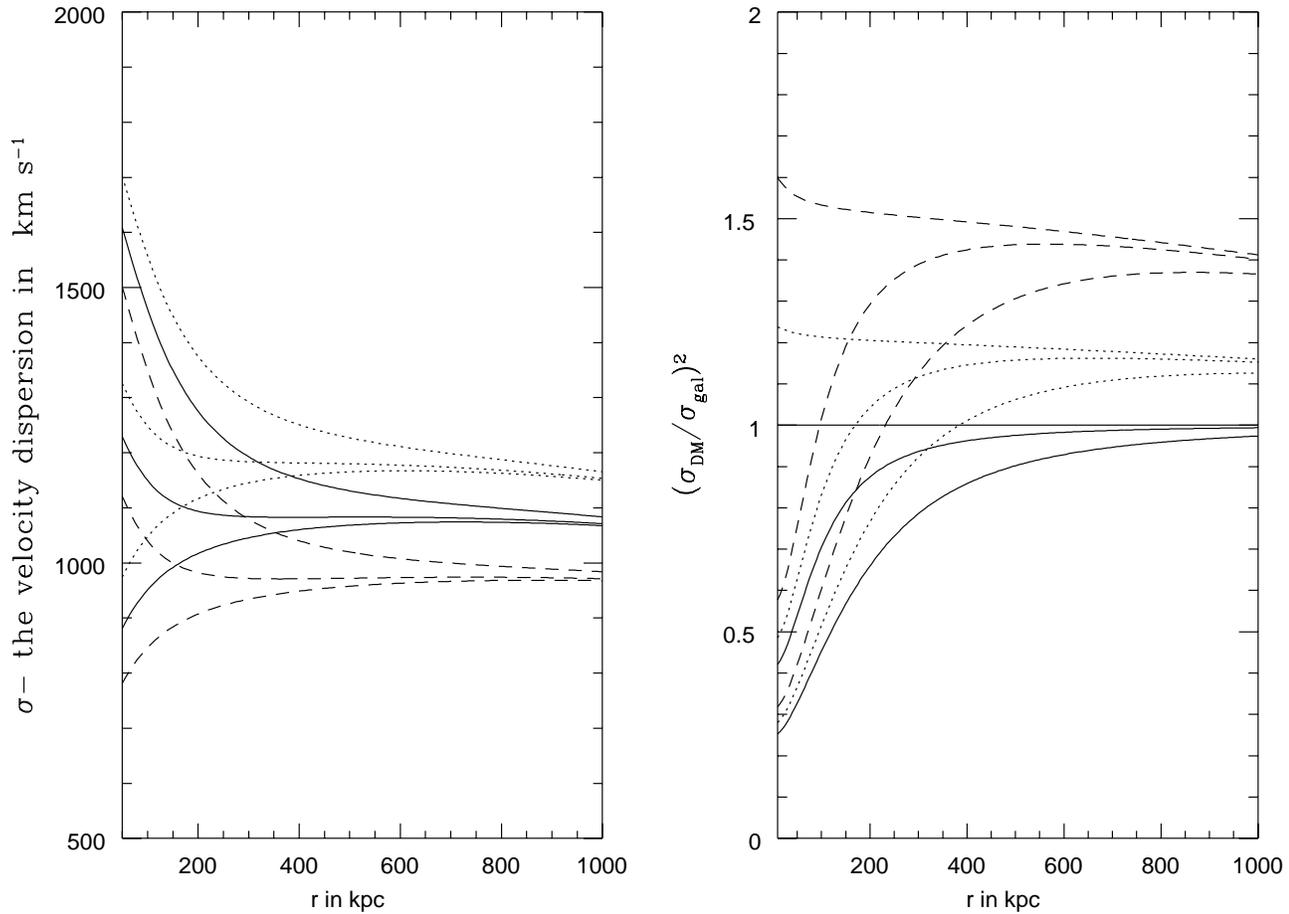,width=1.0\textwidth,angle=270}}
\caption{Left panel: Solutions of the isotropic Jeans equation.  The dotted curves
are for $\alpha$ = 1.0, dashed curves for $\alpha$ = 1.2 and solid curves
for $\alpha$ = 1.5; for each $\alpha$, the smaller the core radius for
the tracer, the smaller the central velocity dispersion (see
text). Right panel:Ratio of the velocity dispersions for various values of
$\alpha$ and core radius $r_{0}$: solid curves - $\alpha_{DM}$/$\alpha_{gal}$ = 1.0,
$r_g$ = 50, 125 and 250 kpc; dotted curves - $\alpha_{DM}$/$\alpha_{gal}$ = 1/1.2, 
$r_g$ = 50, 125 and 250 kpc; dashed curves - $\alpha_{DM}$/$\alpha_{gal}$ = 1/1.5 for 
$r_g$ = 50, 125 and 250 kpc}
\end{figure*}

\subsection{Anisotropic Jeans equation}

The Jeans equation is a mathematical statement of detailed pressure 
balance for an equilibrium stellar system. The dynamical evolution of
clusters in N-body simulations has been studied using the distribution
function formalism \cite{natarajan96b}. This analysis indicates that
clusters evolve from one quasi-equilibrium state to another. A cluster
in a quasi-equilibrium configuration is found to be virialized and has
a smooth potential, which is traced by galaxies with one of the
following orbital structures :
\begin{enumerate}
\item $\beta = 0$, isotropic orbits 

\item $0 < \beta \leq 1$, orbits are mostly radial

\item$\beta < 0$, when the orbits are primarily transverse. 
\end{enumerate}
While $\beta$ has no lower bound, it is strictly required to be less
than 1 for any physically admissible solutions for the velocity
dispersion. 
In this context, it is instructive to examine and compare with studies
of the formation and evolution of elliptical galaxies \cite{hjorth91}. 
It has been shown that the observed uniformity in the properties of
elliptical galaxies can arise from either of two sets of
initial conditions: dissipationless cold collapse or a `warm collapse' 
(or merger) with dissipation. The predicted evolution to the final state with a
deep potential and significant radial anisotropy arises from the
relaxation brought about by global potential fluctuations rather than
two-body encounters (\citeN{aguilar90} and \citeN{londrillo91}). 
Therefore, radial anisotropy can arise naturally in most models as a
consequence of relaxation, and as demonstrated by \citeN{gerhard93}
the line-of-sight velocity profiles being more sensitive to $\beta$
and less so to the potential or to the stellar number density profile
provide a probe of the kinematics of the core. Conversely, for a
galaxy cluster, the initial collapse conditions are 
different and additionally, many physical processes that can effect
energy exchange are active and do occur in the dense
core region. For a cluster core that has virialized, we expect the
orbits in the core to reflect the efficiency of the energy exchange 
mechanisms, while outside the core region, we expect and do find that 
the orbits are largely radial, $0 < \beta \leq 1$ \cite{natarajan96b}.

\section{Proposed Approach}

\subsection{\bf THE MATHEMATICAL FORMALISM}

In our approach, we solve the full Jeans equation for the
velocity anisotropy parameter $\beta$ and for the radial component of
the velocity dispersion $\sigma_{r}^{2}$, using the projected mass profile for
the cluster as constructed independently from gravitational lensing.

From the observed projected galaxy positions, we fit to
get a surface number density profile $\Sigma_{g}(r)$ and use the Abel
integral inversion to extract the three-dimensional density profile
$\nu_{g}(r)$. The key assumption made in the
analysis below is that of spherical symmetry.  
Starting with the full Jeans equation,
\begin{equation}
\frac{d \,(\nu_{g}\,\sigma_{r}^{2})}{dr} + \frac {2 \beta(r) \nu_{g} \ \sigma_{r}^{2}}
{r}  =  - \frac {\ G \ M_{\rm tot}(r) \nu_{g}}{r^{2}};
\end{equation}
where $\nu_{g}(r)$ is three-dimensional galaxy density profile,
$\sigma_{r}^{2}(r)$ the radial velocity dispersion of the galaxies,
$\beta(r)$ is the velocity anisotropy and  $M_{\rm tot} (r)$ is the
distribution of total mass (most accurately determined from
gravitational lensing).

In addition, we have the equation that defines the observed line-of-sight 
velocity dispersion profile $\sigma_{los} (R)$,

\begin{eqnarray}
\frac{1}{2} \, [\,\Sigma_{g}(R) \, \sigma_{los}^{2}(R) ] \, = 
\, \int_{R}^{\infty} \frac {r \nu_{g}(r) \sigma_{r}^{2}(r)\, dr}{\sqrt{(r^{2} - R^{2})}}  \nonumber \\ 
\,\,\,\,\,\,\, - R^{2} \int_{R}^{\infty} 
        \frac {\beta (r) \sigma_{r}^{2}(r) \nu_{g} (r)\, dr}{ r \sqrt{(r^{2} - R^{2})}}.
\end{eqnarray} 
We need to solve these two integro-differential
equations numerically for $\sigma_{r}^{2}$ and $\beta (r)$. For an individual
galaxy with an assumed mass profile, these coupled equations have been
solved by \citeN{binney82a} and \citeN{bicknell89}.

We truncate the integration at a large, finite truncation radius
$R_{t}$, defined strictly to be the radius at which both $\Sigma_{g}
(R_{t})$ and $\rho_{g} (R_{t})$ tend to zero.  Substituting the
expression for $\beta$ from equation (22) we have,

\begin{eqnarray}
\frac{1}{2}[\,\, \Sigma_{g}(R) \ \sigma_{los}^{2}(R) - \,{R^2}
\,{\int_{R}^{R_{t}}}{\frac {G M_{\rm tot} (r) \nu_{g}}{r^{2} {\sqrt{{r^2} - {R^{2}}}}}} \,dr ] \nonumber \\ = \,\, \,\, {\int_{R}^{R_{t}}}{\frac {r \nu_{g} \sigma_{r}^{2} dr}{{\sqrt{{r^2} - {R^{2}}}}}} + \nonumber \\  {\frac {R^{2}}{2}}{\int_{R}^{R_{t}}}{\frac {d (\nu_{g} \sigma_{r}^{2})}{dr}}{\frac {dr}{{\sqrt{{r^2} - {R^{2}}}}}}.
\end{eqnarray}
Integrating the first term on the right hand side by parts and substituting back we have,
\begin{eqnarray}
\frac{1}{2}[\Sigma_{g}(R) \sigma_{los}^{2}(R) - {R^{2}
{\int_{R}^{R_{t}}} \frac {G M_{\rm tot} (r) \nu_{g}}
{r^{2} {\sqrt{{r^2} - {R^{2}}}}}} dr]\nonumber \\ = \int_{R}^{R_{t}} \frac {(\frac {3 R^{2}}{2} - r^{2})}{{\sqrt{{r^2} - {R^{2}}}}}\frac { d ( \nu_{g} \sigma_{r}^{2})}{dr} \,dr.
\end{eqnarray}
The equation can be further simplified and reduced after some algebra
(for details see \citeNP{bicknell89}) to the following
integrals,
\begin{equation}
\nu_{g} (r) \sigma_{r}^{2} = I_{1} (r) - I_{2} (r) + I_{3} (r) - I_{4} (r),
\end{equation}
\begin{equation}
I_{1} (r) = {\frac {1}{3}}{\int_{r}^{R_t}}{\frac {G M_{\rm tot} (r)\,{\nu_{g}}}{r^{2}}}\, dr,
\end{equation}
\begin{equation}
I_{2} (r) = {\frac {-2}{3 r^{3}}}\int_0^r G M_{\rm tot} (r) \,\nu_g\, r \, dr,
\end{equation}
\begin{equation}
I_{3} (r) = {\frac {1}{r^3}}{\int_{0}^{r}}R \,\,\Sigma_{g} (R) \sigma_{los}^{2} (R)\, dR,
\end{equation}
\begin{eqnarray}
I_{4} (r) = {\frac {2}{\pi r^3}}{\int_{r}^{R_t}}R \,\, \Sigma_{g} (R) \, \sigma_{los}^{2} (R) [ \,
{\frac {r}{\sqrt{{R^{2}} - {r^{2}}}}} \nonumber \\ \, - \, sin^{-1} {\frac {r}{R}} \,]\, dR.
\end{eqnarray}
It is to be noted here that the explicit dependence on the mass profile and 
the observed line-of-sight velocity dispersion profile separate. All the 
above integrals are well-behaved with the exception of $I_{4}$, which 
has an integrable singularity which can be taken care
of easily via a simple transformation of variables. Computing these
integrals is nevertheless tricky 
as the final profile for $\sigma_{r}^{2} (r)$ is sensitive to the precise 
asymptotic behavior of all the four terms.
\begin{figure*}
\centerline{\psfig{figure= 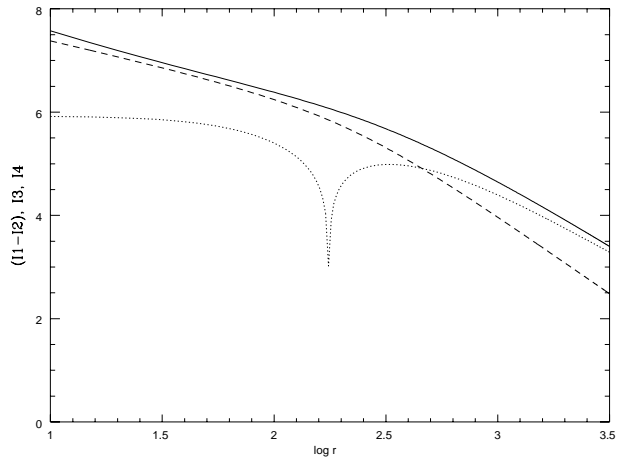,width=0.5\textwidth,angle=270}}
\caption{The  Integrals  - (I1-I2) - dotted curve, I3 - dashed curve, I4
- solid curve, computed for A2218.}
\end{figure*}
The numerical solution for $\sigma_{r}^{2} (r)$ is then substituted back into the
Jeans equation to obtain $\beta$,
\begin{equation}
\beta (r) = -{\frac {r}{2 \nu_{g} \sigma_{r}^{2}}}\,[ \, {\frac {\,G
M_{\rm tot} (r)\, \nu_{g}}{r^{2}}}
+ {\frac {d}{dr}}( \rho_{g} \sigma_{r}^{2})\,\,].
\end{equation}
The variation of $\beta$ with radius can be understood physically in
terms of the relative importance of the mass term and the `galaxy
pressure' gradient term. Rewriting the above equation as follows,
\begin{equation}
\beta (r) = -{1\over 2}\,[\,{v_{c}^2(r)\over {\sigma_{r}^2(r)}}\,+\,{{d\,\ln\,
\nu_{g}\,\sigma_{r}^{2}}\over {d\,\ln\, r}}\,]
\end{equation}
we find that the sign of $\beta$ depends crucially on the
asymptotic behavior of the mass model at large $r$, and specifically
for $\rho_{\rm tot}$ ranges between ${r^{-2}}$ and ${r^{-3}}$, it is
found to be fairly insensitive to
the slope of the assumed galaxy density profile.  The sensitivity of
the sign and magnitude of $\beta$ to the slope of the mass profile
enables its use as a discriminant between the various mass models.

\section{Results for various mass profiles}

We consider several physically motivated fiducial density profiles for
the total mass and in what follows, we examine both galaxy
distribution profiles described in equation (1) (PROFILE A) and 
equation (2) (PROFILE B). The asymptotic slope of the
density profile is defined to be $\gamma$. All the mass profiles are 
normalized to have the same total projected mass enclosed within the 
Einstein radius
$[M_{arc}(r_{E}\,=\,r_{arc})\,=\,(5\pm0.1)\times\,10^{13}\,M_\odot]$, 
as calibrated from strong cluster lensing in A2218. The density profiles
and mass models studied are:

{\flushleft{\underline{\bf MODEL I :}}}

\begin{equation}
\rho(r) = \frac {\rho_{0}}{r {(r^{2} +r_{0}^{2})^{\alpha}}}\,\,\,;\,\gamma\,=\,2\alpha + 1,
\end{equation}
\begin{eqnarray}                
M(r) \,=\,2 \pi \rho_{0} \,\, \rm \ln\, {(r^{2} + r_{0}^{2})} \,\,\,;\,\,\, \alpha = 1.
\end{eqnarray}

{\flushleft{\underline{\bf MODEL II :}}}

\begin{equation}
\rho(r) = \frac {\rho_{0}}{(r^{2} + r_{0}^{2})^{\alpha}}\,\,\,;\,\gamma\,=\,2\alpha, 
\end{equation}
\begin{equation}
M(r) \,=\, 4 \pi \rho_{0} r_{0}\, [ (\frac{r}{r_{0}})  -  tan^{-1} (\frac{r}{r_{0}})\, ] \,\,\,;\,\,\,\, \alpha = 1.
\end{equation}

{\flushleft{\underline{\bf MODEL III :}}}

\begin{equation}
\rho(r) = \frac {\rho_{0}}{ r (r + r_{0})^{\alpha}}\,\,\,;\,\gamma\,=\,\alpha + 1,
\end{equation}
\begin{equation}
M(r) \,=\, 4 \pi \rho_{0} r_{0}\, [\, \frac{r}{r_{0}} - \rm \ln \, ( r + r_{0})\, ]  \,\,\,;\,\,\,\, \alpha  = 1.
\end{equation}  

\subsection{Dependence on the slope $\gamma$}

We analyse here the results for the various mass models with a
specified asymptotic slope $\gamma$, assuming a core radius of  
$r_{g}\,=\,250\,$kpc for the galaxy distribution of PROFILE A 
with $r_{0}\,=\,60\,$kpc for the dark matter.
\begin{figure*}
\centerline{\psfig{figure= 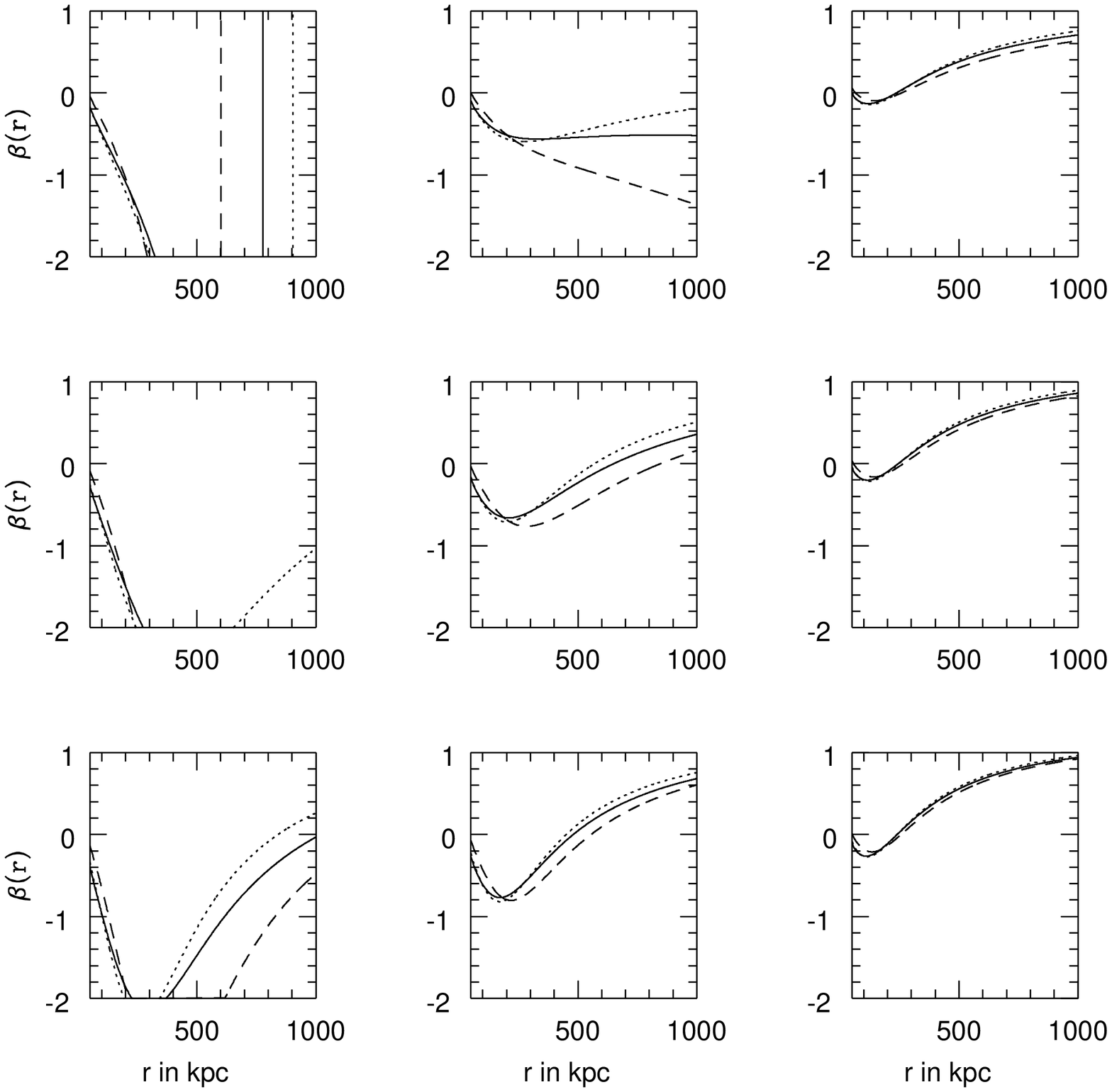,height=0.6\textheight,width=1.0\textwidth}}
\caption{Top Panel: $\beta(r)$  for PROFILE A with $r_{g}$ = 250 kpc,
the mass models with asymptotic slope $\gamma$ = -2.0;  
solid curve - Model I, dotted curve - Model II, dashed curve - Model
III, for $\sigma_{los}$ = 800, 1000 and 1400 km s$^{-1}$
respectively. Centre Panel: $\beta(r)$  For the mass models with
asymptotic slope  $\gamma$ = -2.5;  solid curve - Model I, dotted
curve - Model II, dashed curve - Model III, for
$\sigma_{los}$ = 800, 1000 and 1400 km s$^{-1}$ respectively. Bottom
Panel: $\beta(r)$  For the mass models with asymptotic slope  $\gamma$
= -3.0;  solid curve - Model I, dotted curve - Model II, dashed curve
- Model III, for $\sigma_{los}$ = 800, 1000 and 1400 km s$^{-1}$ 
respectively.}
\end{figure*}

\begin{figure*}
\centerline{\psfig{figure= 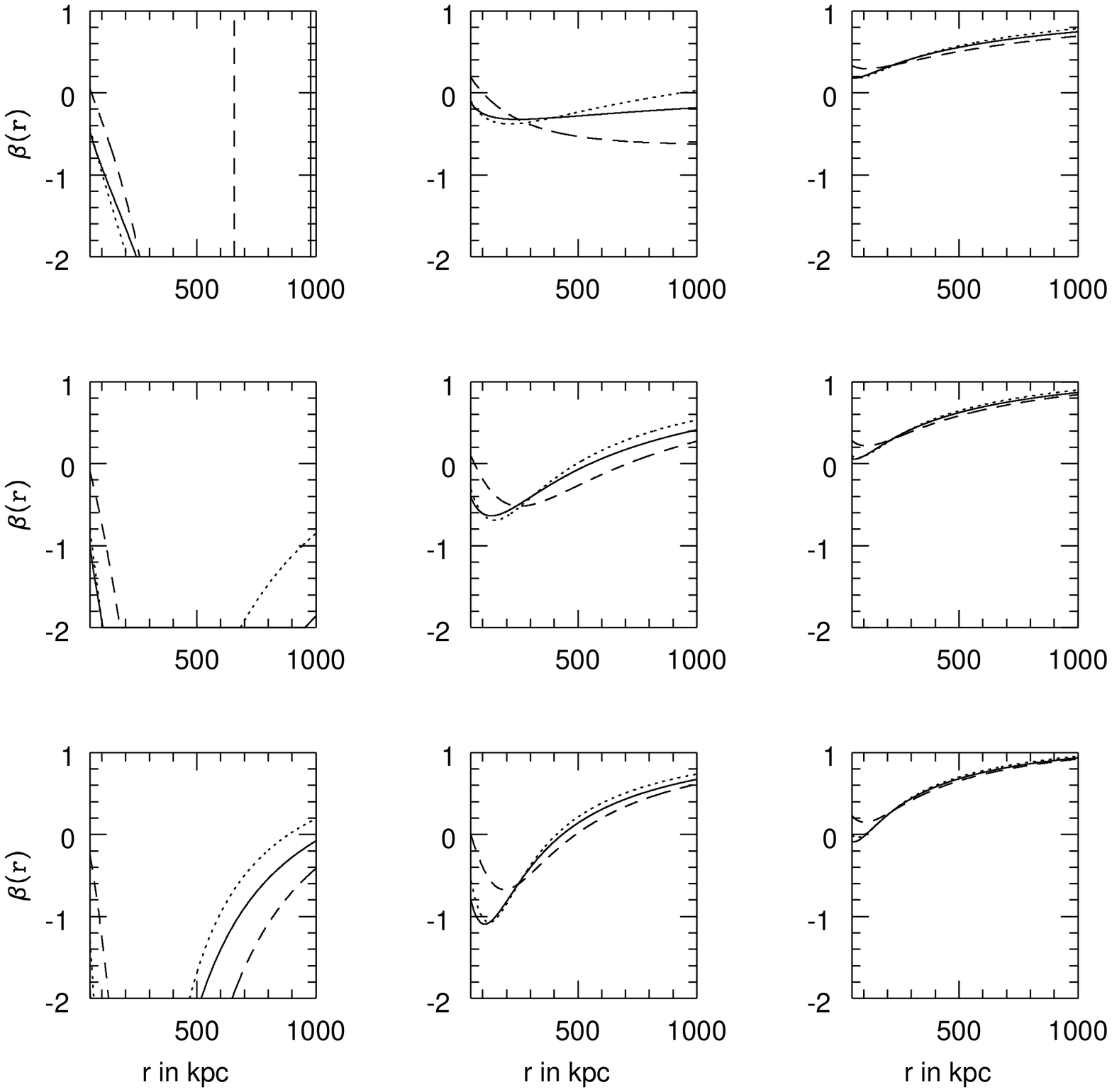,height=0.6\textheight,width=1.0\textwidth}}
\caption{Top Panel: $\beta(r)$  for PROFILE B with $s$ = 200 kpc,
  the mass models with asymptotic slope $\gamma$ = -2.0;  solid curve
  - Model I, dotted curve - Model II, dashed curve - Model III, for 
$\sigma_{los}$ = 800, 1000 and 1400 km s$^{-1}$ respectively. 
Centre Panel: $\beta(r)$  For the mass models with asymptotic slope  
$\gamma$ = -2.5;  solid curve -
Model I, dotted curve - Model II, dashed curve - Model III, for
$\sigma_{los}$ = 800, 1000 and 1400 km s$^{-1}$ respectively. Bottom
Panel: $\beta(r)$  For the mass models with asymptotic slope  $\gamma$
= -3.0;  solid curve - Model I, dotted curve - Model II, dashed curve
- Model III, for $\sigma_{los}$ = 800, 1000 and 1400 km s$^{-1}$ 
respectively.}
\end{figure*}

\begin{enumerate}
\item For $\gamma\,=\,-2.0$ (Fig. 3 - top panel), and a range of input values of the line of sight
velocity dispersion assumed to be constant $(\sigma_{los}\,=\,800,\,1000,\,and\,
1400\,$km s$^{-1}$); we obtain unphysical solutions
$(\sigma_{r}(r)\,<\,0)$ for the lowest $\sigma_{los}$ for all
the 3 models. On increasing $\sigma_{los}$ to $1000\,$km s$^{-1}$, the orbits are primarily 
transverse in the core progressing to more radial ones in the outer
parts, so that $\beta\,<\,0$. For the highest $\sigma_{los}$ we find
evidence for a small core region with mixed orbits, but with primarily
radial orbits outside 200 kpc for all 3 models.
\item For $\gamma\,=\,-2.5$ (Fig. 3 - middle panel), unphysical solutions 
are obtained for the lower $\sigma_{los}$ value for all models, but for 
$\sigma_{los}\,=\,1000\,\,and\,1400\,$km s$^{-1}$ we do find
physically admissible solutions. All 3 mass models have a finite core
with mixed orbits leading on to largely radial orbits outside. Model I
has the largest mixed region (of the order of 700 kpc) while Models II
and III have smaller mixed regions which are of the order of 500 kpc. 
The highest value of $\sigma_{los}$ produces primarily radial orbits
from the centre outward. 
\item For $\gamma\,=\,-3.0$ (Fig. 3 - bottom panel), and the lowest value of $\sigma_{los}$, we
obtain unphysical solutions, but as $\sigma_{los}$ is increased there
is evidence for a core with transverse orbits. 
\end{enumerate}

Assuming the galaxy distribution to be of the form of PROFILE B with a
scale radius $s\,=\,200\,$kpc and $r_{0}\,=\,60\,$kpc for the dark
matter, we find the following trends:

\begin{enumerate}
\item For $\gamma\,=\,-2.0$ (Fig. 4 - top panel), and the same range of input
  values of the line of sight
velocity dispersion $(\sigma_{los}\,=\,800,\,1000,\,and\,
1400\,$km s$^{-1}$), we obtain unphysical solutions in the core
 region for the lowest $\sigma_{los}$ for all
the 3 models. On increasing $\sigma_{los}$ to $1000\,$km s$^{-1}$, the
orbits tend to be tranverse so that $\beta\,<\,0$. 
For the highest $\sigma_{los}$ we find primarily radial orbits for all 3 models.
\item For $\gamma\,=\,-2.5$ (Fig. 4 - middle panel), unphysical solutions 
are obtained for the lower $\sigma_{los}$ value for all models, but for 
$\sigma_{los}\,=\,1000\,\,$ we find that all 3 mass models have a
finite core (of the order of 500 kpc) with mixed orbits leading on to 
largely radial orbits outside. The highest value of $\sigma_{los}$
produces primarily radial orbits right from the centre outward. 
\item For $\gamma\,=\,-3.0$ (Fig. 4 - bottom panel), once again for
  the lowest value of $\sigma_{los}$, we obtain unphysical solutions, 
but as $\sigma_{los}$ is increased there is evidence for a core with 
tangential anisotropy.
\end{enumerate}
Both PROFILES A and B require high values of the line-of-sight velocity dispersion
to produce physically meaningful solutions. The trends above seem to be
qualitatively consistent with the physical picture of ongoing
isotropization and regularization in the core for all the 3 fiducial
mass models considered.

\subsection{Dependence on the circular velocity}

An important parameter for the dynamics of the galaxies in the global
cluster potential is the
circular velocity, $v_{c}$ (Fig. 4), which
measures the change in slope of the mass profile.
For a given mass model with asymptotic slope $\gamma$,
increasing the core radius increases the circular velocity for all
three  models (where these were normalized to have the same projected mass 
within the radius of the arc). 

\begin{figure*}
\centerline{\psfig{figure= 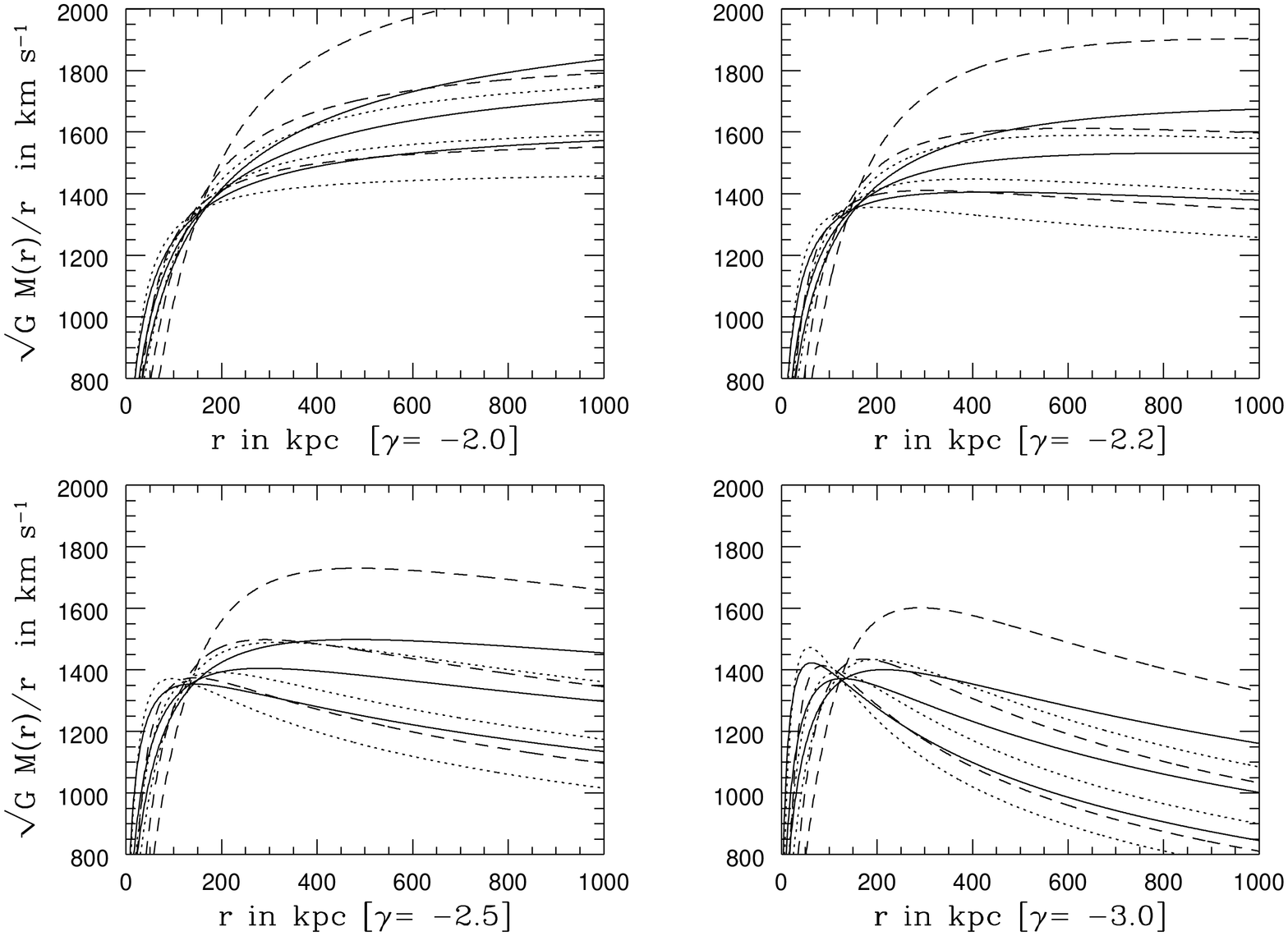,width=1.0\textwidth,angle=270}}
\caption{Asymptotic behavior of the fiducial mass models - varying the
core radius $r_0\,=\,30,\,60,\,100$ kpc; solid curves - Model I,
dotted curves - Model II, dashed curves - Model III}
\end{figure*}
Comparing different mass models that have the same 
asymptotic value of $v_{c}$ (but different $r_0$ and $\gamma$), we find that the
velocity structure of the core and anisotropy profiles are fairly
similar. The qualitative behavior of $\beta$ for fixed asymptotic $v_c$ depends 
strongly on $\sigma_{los}$; with increasing $\sigma_{los}$ we find 
preferentially radial orbits. For a fixed $\sigma_{los}$, increasing
the circular velocity  increases the size of the mixed core region while
lowering the value of $\beta$ at large radius.
For both high and low circular velocities, and $-3.0\,\leq\,\gamma \,\leq\,-2.0$;
low line-of-sight velocity dispersion models $\sigma_{los}\,<\,1000\,$
km s$^{-1}$ are ruled out purely from the dynamical point of view.

\subsection{Dependence on the central density profile} 

The best probe of the shape of the density profile at the very centre
comes from the observed velocity dispersion of the stars in the cD
halo \cite{jme95}. 
For the fiducial mass models of Section 5, we solve for the
line-of-sight velocity dispersion of the cD
halo stars using the isotropic Jeans equation (neglecting the
contribution of the mass of the cD galaxy to the total mass of the cluster):
\begin{equation}
{\frac {d \, (\rho_{cD} \, \sigma^{2}_{*} \,)}{ d r}}= - {\frac {G
M_{\rm tot}(r) \rho_{cD}}{r^{2}}}.
\end{equation}
Assuming a scaling of $\rho_{cD}\,\propto\,r^{-\delta}$, and
$\rho_{tot}\,\propto\,r^{-\gamma}$, close to the centre we obtain: 
\begin{equation}
\sigma_{*}\,=\,A\,\frac{r^{2-\gamma}}{(\delta+\gamma-2)} + const.
\end{equation}
\begin{figure*}
\centerline{\psfig{figure= 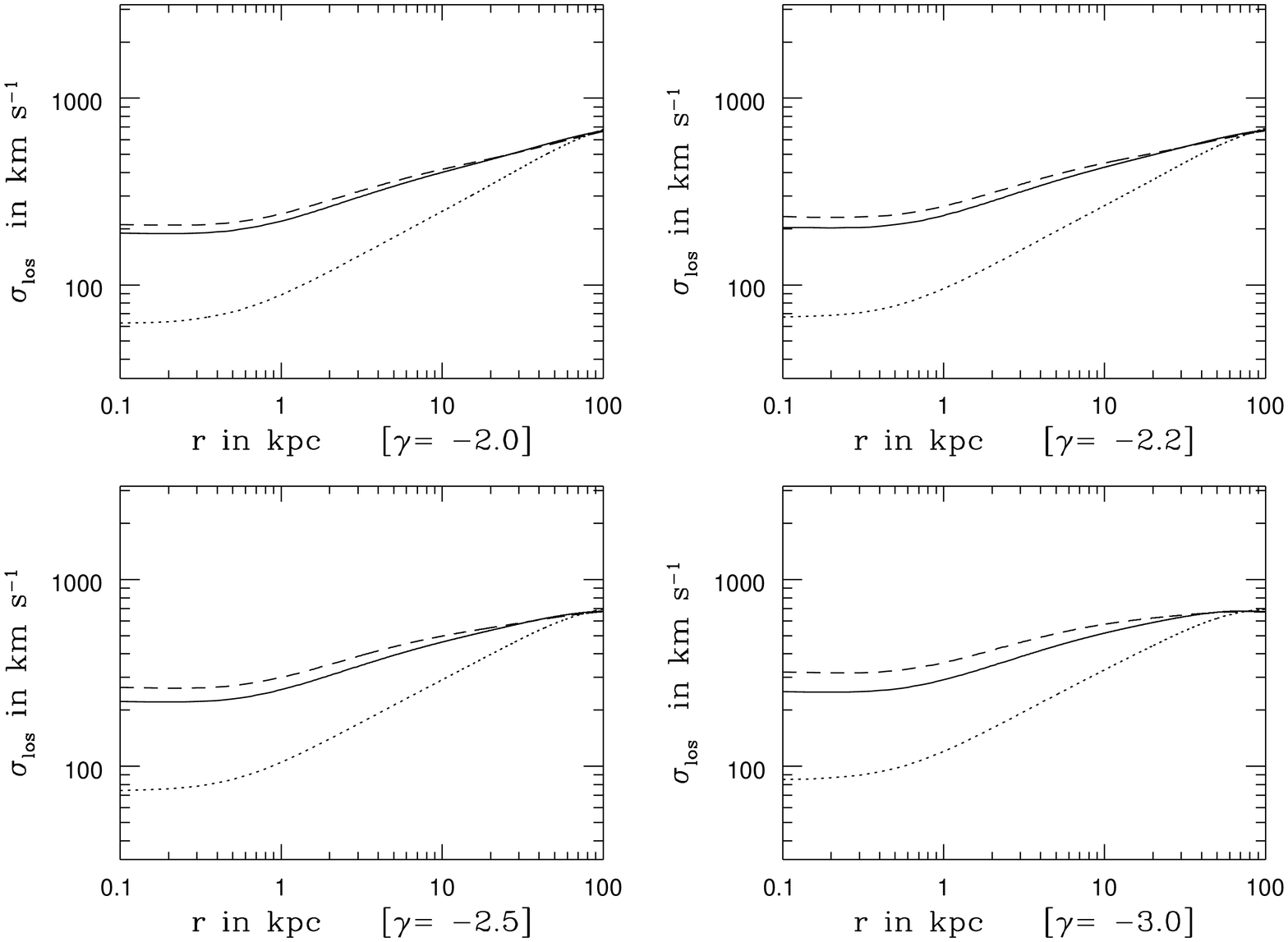,width=1.0\textwidth,angle=270}}
\caption{Computed line-of-sight velocity dispersion for the cD halo stars:
solid curve - Model I, dotted curve - Model II, dashed curve - Model III. }
\end{figure*}
Therefore, for $0\,\leq\,\gamma\,<\,2$ and $\delta+\gamma\,>\,2$, we
expect the velocity dispersion of the stars to rise. The cD profile
from section 2.3.4 is used with a core radius
$r_{1}\,=\,0.05\,$kpc and $r_{cut}\,=\,35.0\,$kpc.
The three models studied in the previous section, predict profiles 
(Fig. 5) with low central values for $\sigma_{*}$ that 
rise steeply with radius. Models I and III have central values $\sim$
200 - 400 km s$^{-1}$, varying with $\gamma$, such that the steeper the total mass
profile the higher the central value. For the stars, $\sigma_{los}$ rises to 
700 km s$^{-1}$ at $r\,=\,100\,$kpc, which is  consistent with the
measurements of the cD galaxy IC1011 in A2029 by \citeN{dressler79} and
\citeN{fisher95}. Model II under-predicts the central value and is
qualitatively incompatible with the data.  

\subsection{Summary of the important parameters}

In the above analysis, there are several parameters to be kept track
of in order to interpret the results for the computed velocity
structure of the cluster core, namely the total mass distribution and the
galaxy positions and velocities. 

The mass profile for a given fiducial model is specified by three
parameters, the central density, the core radius, and the slope. The velocity
dispersion of stars in the cD halo favors mass models with a central
`cusp'.  Lensing constrains (i) the mass within the radius of the
arc (strong regime), (ii) the circular velocity at large radius
(weak-shear), and the generic shape of the profile. It favors compact
cores and rules out steep slopes, 
$\gamma\,<\,-2.5$ for $50\,\leq\,r\,\leq\,600\,$kpc. Our results show
that the dynamics of
the core can be recovered
independent of the fine tuning of individual fiducial models,
given the above constraints from lensing and an observationally 
well-determined line-of-sight velocity dispersion profile.
Although currently limited by the errors in the observationally
determined input quantities, this dynamical approach offers a better understanding
of the physical state of the cluster core and can discriminate
between various mass models for an individual lensing cluster.

\section{Application to A2218}

We apply this technique to the Abell cluster A2218, at a redshift $z =
0.175$, with a mean measured velocity dispersion $\sigma_{mean} \sim
1370^{+160}_{-120} \,$km s$^{-1}$ from 56 cluster members.  
A2218 is a cD cluster with a very peaked mass
distribution and a compact core, hence a large number of
gravitationally distorted arcs and arclets are observed.  The mass
model for this cluster was constructed using ground data
\cite{kneib95a} and refined using HST data by \cite{kneib96}. 
Redshifts of two of the arcs were spectroscopically measured by 
\shortciteN{pello92} and further redshifts of arclets have been determined
by \shortciteN{ebbels96} hence tightly calibrating the mass model.

\subsection{OBSERVATIONAL DATA}

The observational input for the galaxies in A2218 for our analysis
comes primarily from the photometric and spectroscopic survey by
\citeN{leborgne92a}. 
\subsubsection{\bf Line-of-Sight Velocity Dispersion Profile}

In order to construct the line-of-sight velocity dispersion profile
(Fig. 6), the galaxies were binned in $100\,$kpc annuli. The
measurement errors preclude any fitting, and therefore we assumed a constant value for 
$\sigma_{los}$. This simplifying assumption is
the largest source of error for our present analysis, but can be
refined with the availability in the  near future of more tightly
sampled line-of-sight velocity dispersion profiles.

\begin{figure*}
\centerline{\psfig{figure= 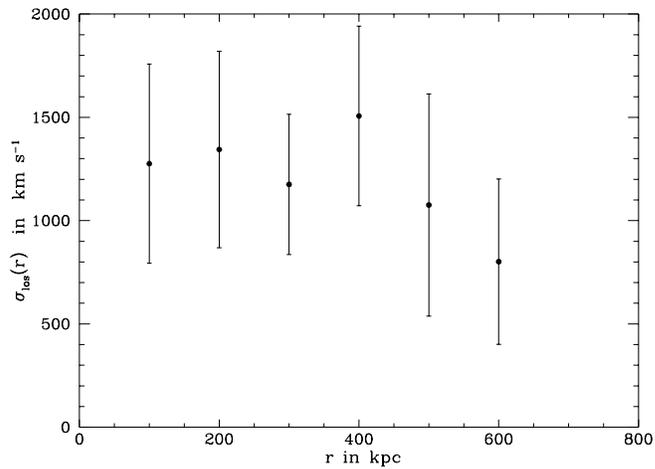,width=0.5
\textwidth,angle=270}}
\caption{The line-of-sight velocity dispersion profile for A2218}
\end{figure*}
\subsubsection{\bf Galaxy density profile from optical data}

The observed surface density of galaxies in A2218 was fitted to a modified Hubble law
profile (PROFILE A) with a core radius $r_{g}\,=\,250\,h_{50}^{-1}\,$kpc,
\begin{equation}
{\Sigma_{g}} (r)\, = \, \frac {\Sigma_{0}}{1 + \frac {r^{2}}{r_{g}^2}},
\end{equation}
and the corresponding 3-D density profile from equation (1), as well
as by the cuspy profile (PROFILE B) with a scale radius
$s\,=\,200\,h_{50}^{-1}\,$kpc,
\begin{equation}
{\Sigma_{g}} (r)\, = \, {\frac {\Sigma_{0}} {{({r \over s})^{0.1}}{(1
      + {r \over s})^{1.9}}}}
\end{equation}
and the corresponding density profile from equation (2).
\subsubsection{\bf Constructing the mass profile}

\begin{figure*}
\centerline{\psfig{figure= 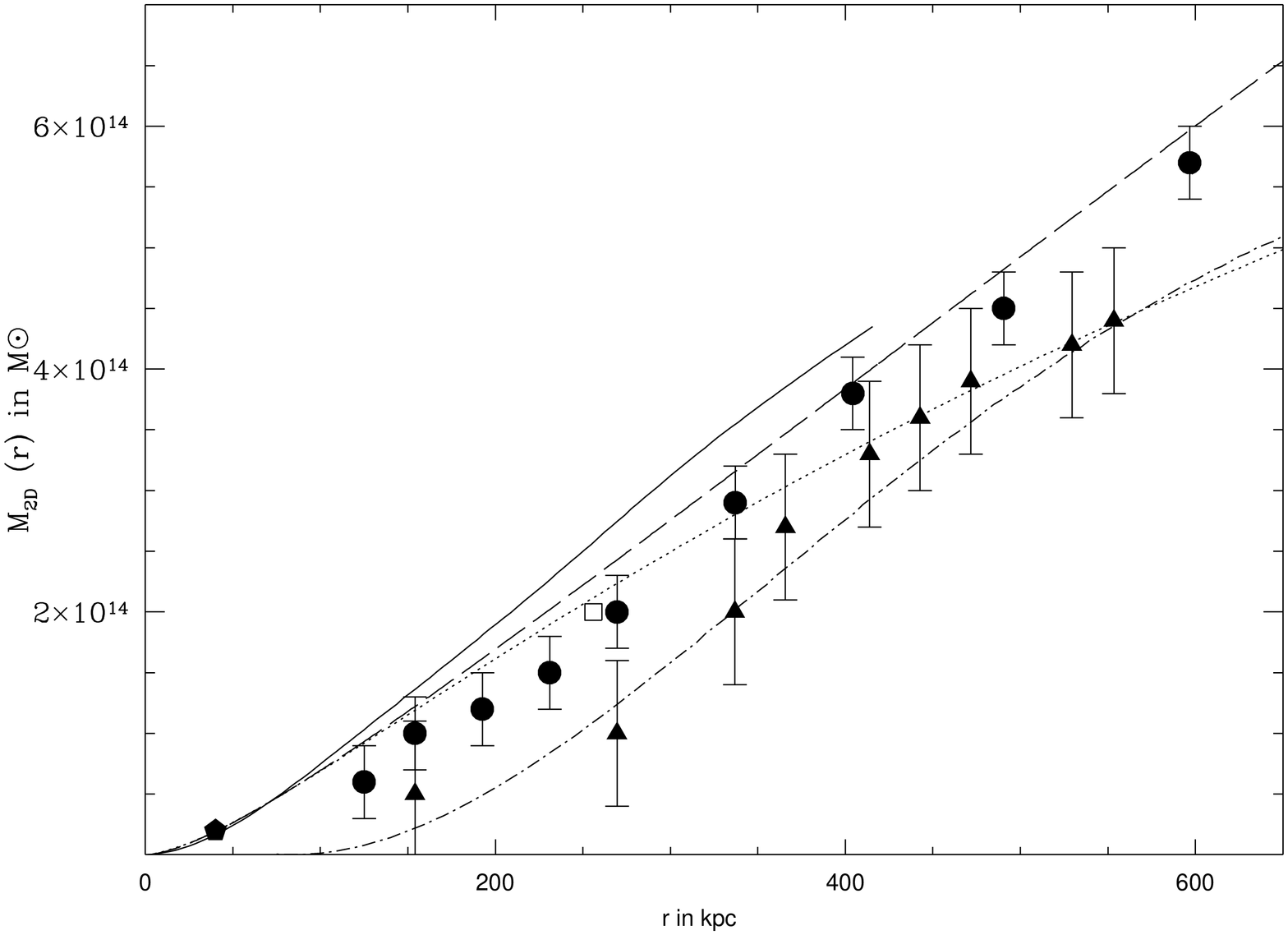,width=1.0\textwidth,angle=270}}
\caption{Projected mass profile: solid triangles - X-ray data, solid
circle - weak lensing mass estimates, filled pentagon - mass from cD, filled square -
mass enclosed within giant arc, solid line - HST mass model, dashed line -
`best fit' composite lens mass model, dot-dashed - fitted X-ray mass
model, dotted line - `best fit' N-body model.}
\end{figure*}

The lensing mass profile for the cluster was constructed from, 
the strong lensing data (arcs, arclets and resolved multiple images) from ground-based 
observations and the HST image by \shortcite{kneib96}, and the
weak-lensing mass map published by \citeN{squires95}. 

A2218 is best fit by the following functional form of MODEL III with
$\alpha\,=\,1.0$ (see Section 5),
\begin{equation}
 M (r) = M_{0}  [ {\frac {r}{r_{0}}} - \rm \ln\,(1 + {\frac{r}{r_{0}}})],
\end{equation}
where we normalize $M_{0}$ to the mass enclosed by the arc at $ r_{arc}\,=\,
78.5\, h_{50}^{-1}\,$kpc and $r_{0}$  the core radius is $60
h_{50}^{-1}\,$kpc. The corresponding three-dimensional density profile
is as below, 
\begin{equation}
\rho (r) = {\frac {\rho_{0} r_{0}^{2}}{ r ( r + r_{0})}},
\end{equation}
with $\rho_{0}\,=\,1\,\cdot\,10^{-22}$ g cm$^{-3}$.
The X-ray mass profile was obtained using the standard deprojection
technique described by \shortciteN{fabian81} to the archival ROSAT HRI map, 
assuming spherical symmetry and hydrostatic equilibrium for the 
intracluster gas. The integrated X-ray luminosity (in the 0.1 -- 2.4 keV band) 
and central temperature of A2218 are measured to be respectively, 
\begin{equation}
L_{x} \,=\, 7\,.\,10^{44} \,\,erg \,\,s^{-1} \,\,\,;\,\, T \,=\,\, 8\, keV,
\end{equation}
in good agreement with the \shortciteN{squires95} results.
The predicted circular velocity is,
\begin{equation}
v_c^{2}(r)\,=\,{G\,M(r)\over r}\,=\,-\frac {k T }{\mu m_{p} }(\,\frac { d\, (\,\rm \ln \,\rho_{gas}\, )}{d\, ln\, r} + \frac {d\,\rm ln T\,}{d\,\rm ln\, r}\,)
\end{equation}
where $M(r)$ is the total mass as inferred from the X-ray analysis,

The mass model from  N-body simulations was also normalized to
the mass enclosed within the Einstein radius, and for consistency with
the observed arcs in A2218 the scale radius
$r_{s}$ (see equation 3) is required to be of the order of
$250\,h_{50}^{-1}\,$kpc \cite{waxman95}.

\subsection{Results}

Using the total mass profile constructed from lensing as described 
above, we solve the equations to obtain solutions (see
Fig. 8) for $\sigma_{r}(r)$, $\sigma_{t}(r)$ (the radial and
transverse velocity dispersion profiles respectively) and the velocity
anisotropy parameter $\beta (r)$.  
\begin{figure*}
\centerline{\psfig{figure= 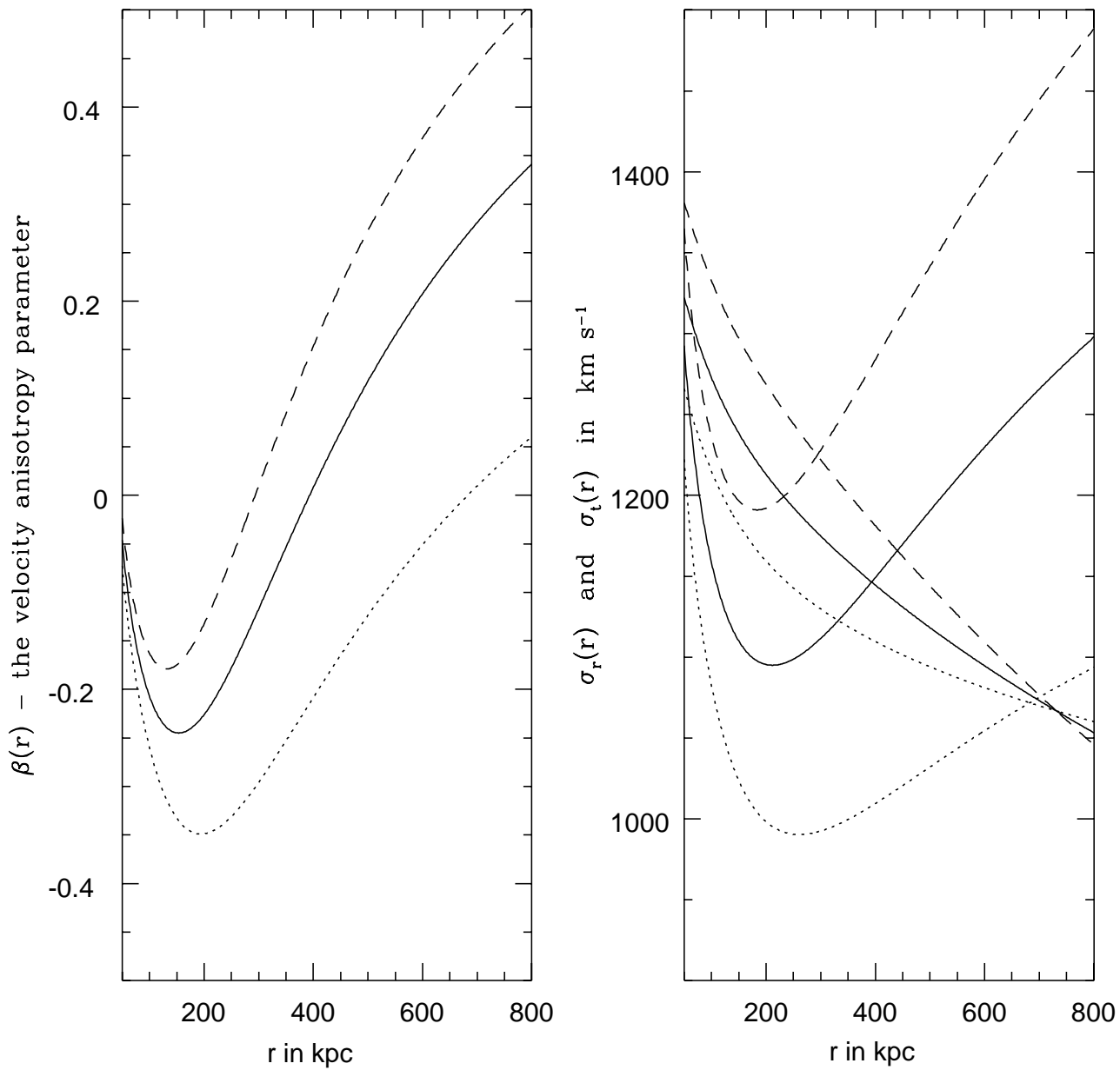,width=1.0\textwidth}}
\caption{The best fit composite mass model for A2218 : The radial and
transverse components of the velocity dispersion and the velocity
anisotropy parameter for $\sigma_{los}\,=\,1100,\,1200,\,1300\,$km
s$^{-1}$ and the galaxy distribution modeled by PROFILE A with
$r_{g}\,=\,250\,h_{50}^{-1}\,$kpc,}
\end{figure*}

\begin{figure*}
\centerline{\psfig{figure= 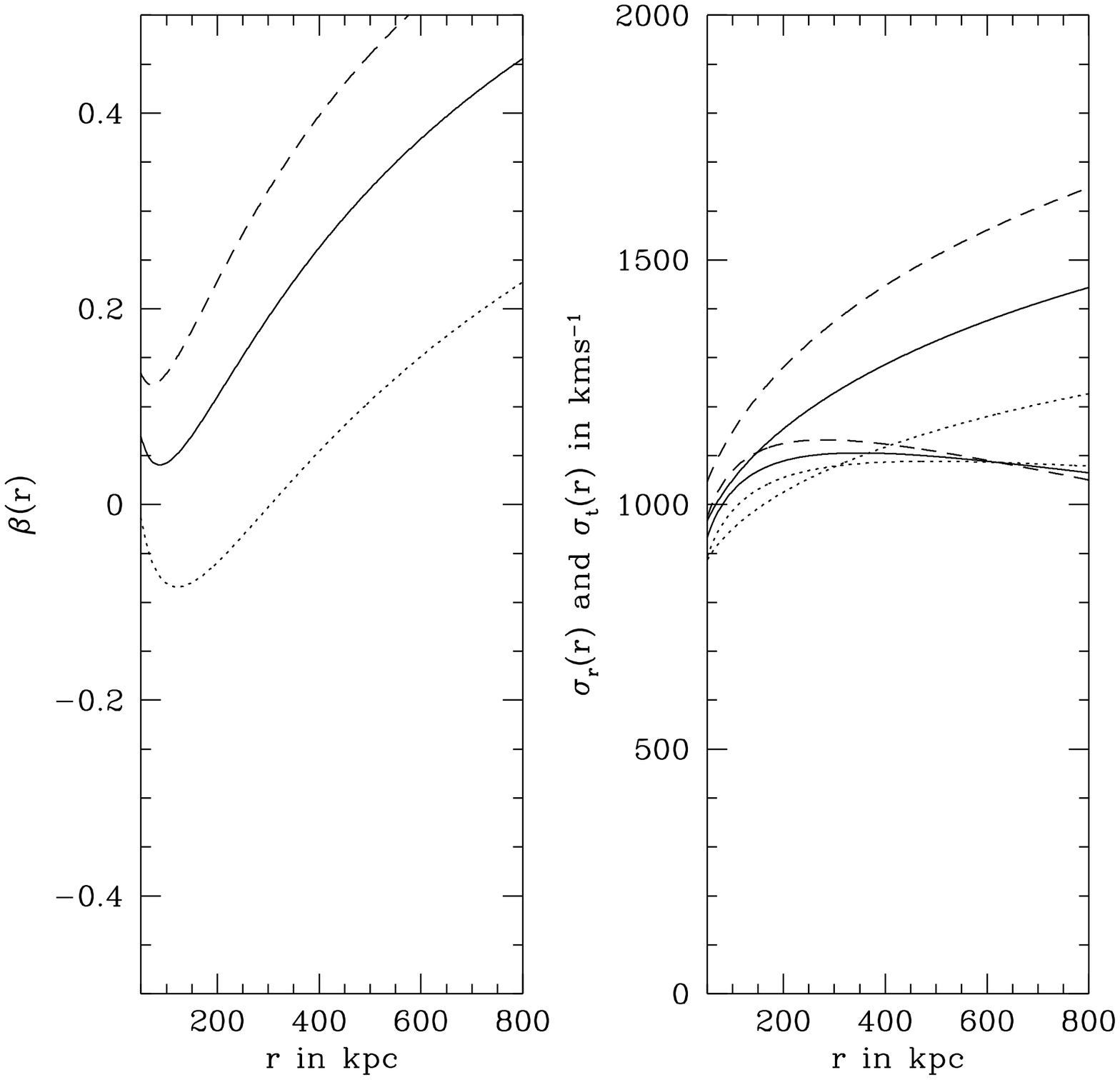,width=1.0\textwidth}}
\caption{The best fit composite mass model for A2218 : The radial and
transverse components of the velocity dispersion and the velocity
anisotropy parameter for $\sigma_{los}\,=\,1100,\,1200,\,1300\,$km
s$^{-1}$ and the galaxy distribution modeled by PROFILE B with
$r_{s}\,=\,200\,h_{50}^{-1}\,$kpc and $\alpha = 0.1$.}
\end{figure*}

The profile was also checked for consistency with the measured stellar
velocity dispersion of the halo stars in the cD galaxy (Fig. 9).
[It is to be noted here that the measured line-of-sight velocity
dispersion profile for A2218 is inconsistent with an isotropic solution.]
\begin{figure*}
\centerline{\psfig{figure= 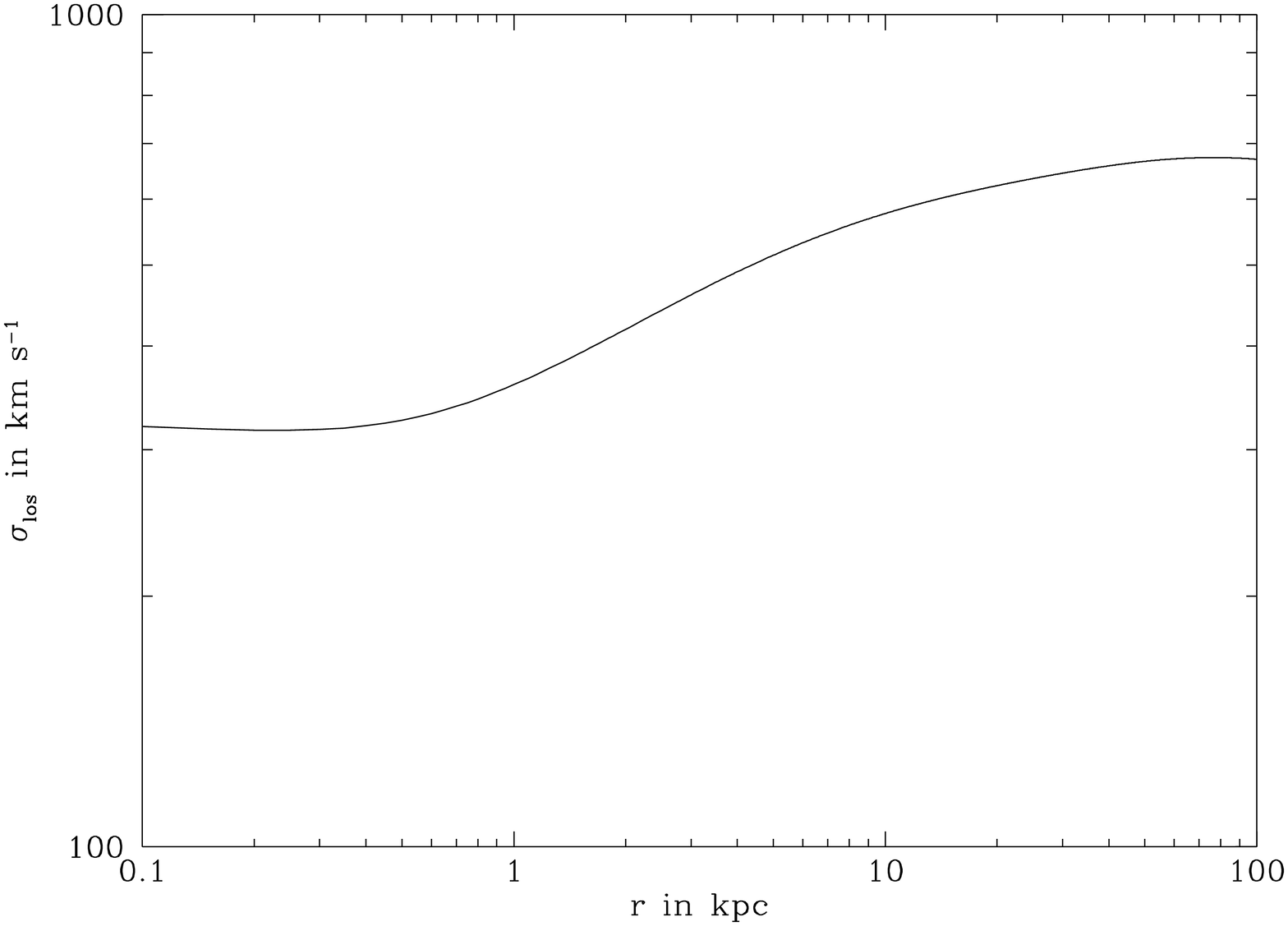,width=0.5\textwidth,angle=270}}
\caption{Computed line-of-sight velocity dispersion profile for cD
halo stars of A2218}
\end{figure*}
We find that the orbits predicted for the best-fit mass model 
in the central regions, isconsistent with the picture of a core not 
in equilibrium, independent of the assumed form for the galaxy number
density distribution.  The precise nature of orbits {\it transverse}
($\beta\,<\,0$) or {\it radial} ($0\,>\,\beta\,>\,1$) depends
on the detailed shape of the line-of-sight velocity dispersion
profile, which is not measured to adequate precision at present.
Both $\sigma_{r}$ and $\sigma_{t}$ fall within the inner $600\,h^{-1}_{50}\,$kpc,
with $\sigma_{r}$ declining more rapidly and then tending to flatten off. From
the slope of $\beta$, the trend with increasing $r$ is that the nature of orbits
tends to being mainly radial at the outskirts,
signaling the existence of a region dominated by infall.
The physical picture that emerges for the description of the dynamical
state of A2218 is one of a dynamically disturbed cluster core.
For lower values of the measured line-of-sight velocity dispersion, we
find a tendency for the predominance of transverse orbits in the central $400\,$kpc (which
is precisely of the order of the distance between the 2 distinct
optical clumps seen in the HST image) and could be interpreted as an 
indication of on-going energy exchange in the core. 
Using the mass model from N-body simulations as the input, we find
that the resultant predictions for $\beta(r)$ agree well with those
calculated for the mass profile reconstructed from lensing. 
For the mass profile from X-ray data for A2218, we obtain qualitative
agreement with the predictions from the lensing mass model. 

\section{Conclusions}

Gravitational lensing provides the `cleanest' way to construct the
total mass profile for a cluster independent of the kinematic details;
additionally, combining strong and weak lensing removes the scaling
ambiguity allowing the calibration of other independent mass models. 
With `good' data for an individual cluster,
the requirements for consistency on the smallest to the largest scales are
stringent enough to constrain the slope of generic density profiles for rich clusters. 
Accurate mass profiles are crucial to settling many important
issues such as the baryon fraction problem and in understanding the
discrepancies and biases arising in the X-ray, lensing and virial mass estimates
for clusters.

In this paper, we have demonstrated that the dynamics and velocity
structure of the core of galaxy clusters can be probed given an
independently inferred total mass profile. The future applications
of our method to study cluster cores are promising, given the prospect
of collecting more spectro-photometric data of galaxies in cluster 
lenses (e.g. \citeNP{yee96}).

With current data, we find strong evidence for the existence
of an anisotropic central region. This is consistent with the picture of on-going
relaxation, wherein anisotropies in the velocity tensor can arise
naturally as a consequence of the initial conditions coupled with evolution. 
Given the range of complex physical processes that operate in cluster
cores that could alter galaxy orbits, for instance; dynamical friction
(dynamical friction in an aspherical cluster can induce and amplify
the velocity anisotropy as demonstrated by \citeN{binney77},
a possible origin for the inferred velocity anisotropy,
specially in the case of A2218), potential fluctuations arising due to
the presence of substructure and the frequent presence of a cD galaxy
at the centre of the cluster potential it is not surprising that the
core is not isotropic.

Distinguishing between the dynamical effects of the various physical  
mechanisms in order to model them satisfactorily, in addition to
requiring from the observations more accurately determined
line-of-sight velocity dispersion profiles for clusters would enable
the application of this technique more effectively. 
Further extension of this analysis to incorporate the
dynamics of the intra-cluster gas with the lensing model
self-consistently, is required in order to understand the possible
role of baryons in the dynamics of cluster cores.

\section*{ACKNOWLEDGEMENTS}

We thank Martin Rees for his support and encouragement during the
course of this work. We acknowledge useful discussions with Richard
Ellis, Bernard Fort, Yannick Mellier, Simon White and Tim de Zeeuw. 
We thank Jens Hjorth and Donald Lynden-Bell for incisive and detailed
comments on the draft, Steve Allen for helping with the data
processing of the HRI archive data and Roser Pell\'o for providing us 
with the A2218 spectro-photometric catalog in computer-readable form. 
We thank the anonymous referee for useful suggestions.
PN acknowledges funding from the Isaac Newton Studentship and Trinity 
College at the University of Cambridge. JPK acknowledge support from 
an EC-HCM fellowship.

%
%

\bibliography{mnrasmnemonic,refs}
\bibliographystyle{mnrasv2}
\end{document}